\documentclass{ptapap}

\author{L.~Moln\'ar}[molnar.laszlo@csfk.mta.hu,Konkoly]
\author{A.~Derekas}[Gothard,Konkoly]
\author{R.~Szab\'o}[Konkoly]
\author{L.~Szabados}[Konkoly]
\author{J.~M.\ Matthews}[UBC]
\author{C.~Cameron}[CBU]
\author{A.~F.~J.~Moffat}[Montreal]
\author{N.~Richardson}[Ritter]
\author{N.~R.~Evans}[SAO]
\affil[Konkoly]{Konkoly Obs., MTA CSFK, Konkoly Thege \'ut 15-17, H-1121 Budapest, Hungary}
\affil[Gothard]{ELTE Gothard Astrophys.\ Obs., H-9704 Szombathely, Szent Imre h.\ \'ut 112, Hungary}
\affil[UBC]{University of British Columbia, 6224 Agricultural Road, Vancouver, Canada}
\affil[CBU]{Cape Breton University, 1250 Grand Lake Road, Sydney, Nova Scotia, Canada}
\affil[Montreal]{Universit\'e de Montr\'eal, C.P. 6128, Succ.\ Centre-Ville, Montr\'eal, Qu\'ebec, Canada}
\affil[Ritter]{Ritter Observatory, The University of Toledo, Toledo, OH, USA}
\affil[SAO]{Harvard-Smithsonian Center for Astrophysics, 60 Garden str., Cambridge, MA, USA}

\title{The MOST view of Cepheids}

\begin{document}

\maketitle

\begin{abstract}

The \textit{MOST} space telescope observed four Cepheid variables so far, all of different subtypes. Here we summarize the results obtained and the possible ways to continue to study these stars.

\end{abstract}

\section{Introduction}
Classical Cepheid stars are radially pulsating supergiants that constitute an important rung in the cosmic distance ladder. The Canadian microsatellite \textit{MOST} observed four Cepheid variables so far. RT Aur and SZ Tau, a fundamental-mode and a first-overtone star, respectively, were studied by \citet{2015MNRAS.446.4008E}. The detailed results about U TrA, a double-mode star, and V473 Lyr, a second-overtone pulsator, will be presented by Moln\'ar et al.\ (in prep.).

\section{RT Aur and SZ Tau}
The first two stars were observed in order to investigate the stability of the pulsation. The observations suggest that SZ Tau shows comparatively stronger O--C variations than RT Aur, suggesting that overtone pulsators are less stable. However, both data sets are short, with less than six cycles covered. Fortunately, SZ Tau will be observable in Campaign 13 of the K2 mission that may provide longer and even more accurate photometry. An interesting question that K2 could answer is whether the low-amplitude $f_X$ mode, seen in many first-overtone stars at frequency ratios $f/f_X \sim 0.62$, is present in SZ Tau too \citep[see, e.g.][]{2016MNRAS.458.3561S}. 

\begin{figure}[h]
\includegraphics[width=\textwidth]{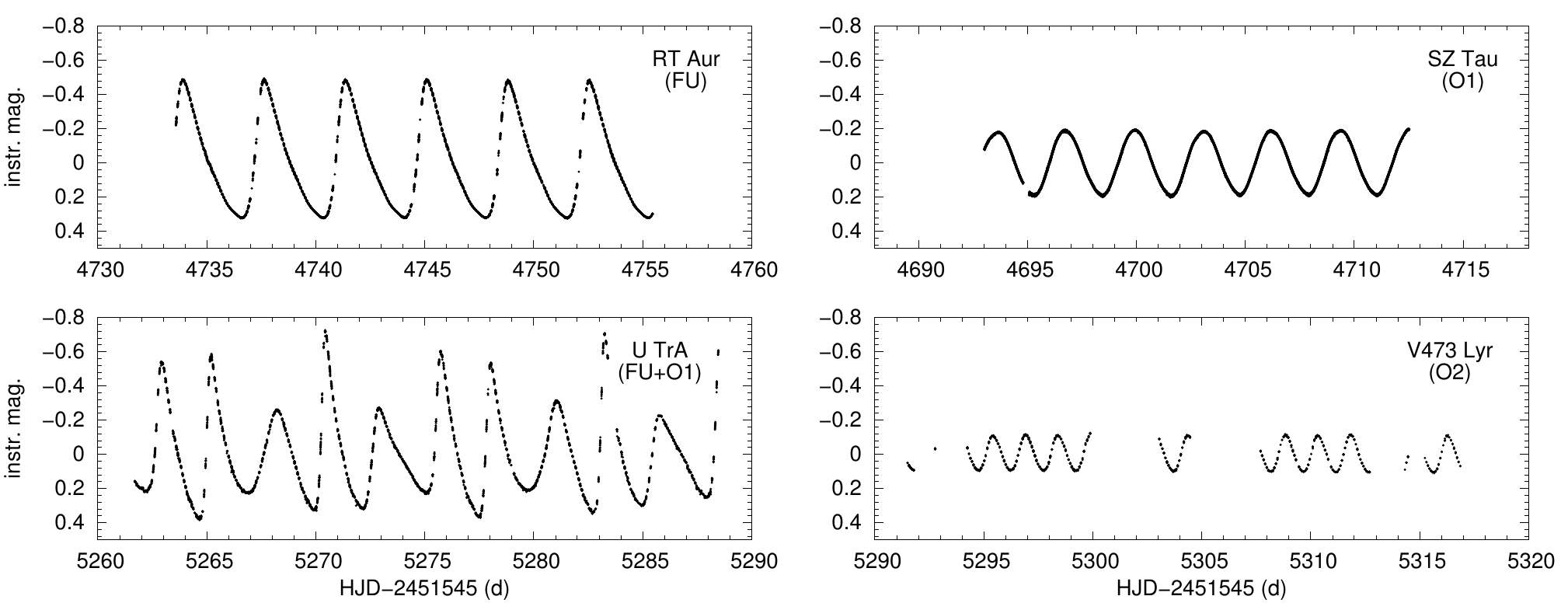}
\caption{The \textit{MOST} light curves of the four Cepheid stars.}
\label{fig:piratevswarming}
\end{figure}

\section{U TrA and V473 Lyr}
The double-mode star U TrA was not a proposed target but it fell into the field-of-view of a primary target. The additional mode $f_X$ was tentatively identified in the star. Moreover, no strong indications of an energy exchange between the modes were identified, contrary to the proposal of \citet{1979ApJ...232..197F}. 

V473 Lyr is one of the few Cepheids with strong Blazhko effect-like modulations \citep{2014MNRAS.442.3222M}. While the \textit{MOST} observations were far too short to cover the 1205-day modulation cycle, they were sufficient to detect alternating higher- and lower-amplitude cycles known as period doubling. The presence of period doubling hints that the modulation could originate from nonlinear mode interactions as suggested for RR Lyrae stars \citep{2011ApJ...731...24B}.

Both U TrA and V473 Lyr will be observable by \textit{TESS}. U TrA will be covered for 27 days, e.g., the same length as the \textit{MOST} data. V473 Lyr could potentially fall into the overlapping area of two sectors, extending the measurements to 54 days. The star will be close to minimum-amplitude phase in 2019, offering an interesting comparison to the \textit{MOST} data taken slightly before maximum-amplitude phase.

\acknowledgements{Based on data from MOST (Microvariability \& Oscillations of STars), which was, at the time the data reported here were collected, a Canadian Space Agency mission operated jointly by Microsatellite Systems Canada Inc.\ (MSCI), and the Universities of Toronto and British Columbia, with support from the University of Vienna. This project has been supported by the K-115709 and PD-116175 grants of the Hungarian National Research, Development and Innovation Office. L.M.\ was supported by the J\'anos Bolyai Research Scholarship. NDR is grateful for postdoctoral support by the University of Toledo and by the Helen Luedtke Brooks Endowed Professorship. AFJM and JMM are grateful for financial aid from NSERC (Canada). AFJM  also acknowledges FQRNT (Qu\'ebec). }

\bibliographystyle{ptapap}
\bibliography{molnar_brite2_proc}

\end{document}